\documentclass[journal]{IEEEtran}

\usepackage{multicol}
\usepackage{multirow}
\usepackage{adjustbox}
\usepackage{hyperref}
\usepackage{makecell}
\usepackage{booktabs}
\usepackage{float}
\usepackage{array, makecell}

\begin{document}
%
\title{Bare Demo of IEEEtran.cls\\ for IEEE Journals}
%
%
%

\author{Michael~Shell,~\IEEEmembership{Member,~IEEE,}
        John~Doe,~\IEEEmembership{Fellow,~OSA,}
        and~Jane~Doe,~\IEEEmembership{Life~Fellow,~IEEE}
\thanks{M. Shell was with the Department
of Electrical and Computer Engineering, Georgia Institute of Technology, Atlanta,
GA, 30332 USA e-mail: (see http://www.michaelshell.org/contact.html).}
\thanks{J. Doe and J. Doe are with Anonymous University.}
\thanks{Manuscript received April 19, 2005; revised August 26, 2015.}}

\title{Fetal-BET: Brain Extraction Tool for Fetal MRI}
\author{Razieh Faghihpirayesh, \IEEEmembership{Student Member, IEEE}, 
Davood Karimi,
Deniz Erdo{\u{g}}mu\c{s}, \IEEEmembership{Senior Member, IEEE},
and Ali Gholipour, \IEEEmembership{Senior Member, IEEE}
%
\thanks{This work has been submitted to the IEEE Transactions on Medical
Imaging for possible publication. Copyright may be transferred without
notice, after which this version may no longer be accessible.}
\thanks{R. Faghihpirayesh is with the Electrical and Computer Engineering
Department, Northeastern University, Boston, MA 02115 USA, and also
with the Radiology Department, Boston Children’s Hospital, and Harvard
Medical School, Boston, MA 02115 USA (e-mail: raziehfaghih@ece.neu.edu).}
\thanks{D. Erdo{\u{g}}mu\c{s} is with the Electrical and Computer Engineering Department, Northeastern University, Boston, MA 02115 USA.}
\thanks{D. Karimi and A. Gholipour are with the Radiology Department, Boston Children’s Hospital, and Harvard Medical School, Boston, MA 02115 USA.}
}

\maketitle

\begin{abstract}
\label{sec:abstract}
 Fetal brain extraction is a necessary first step in most computational fetal brain MRI pipelines. However, it poses significant challenges due to non-standard fetal head positioning, fetal movements during examination, and vastly heterogeneous appearance of the developing fetal brain and the neighboring fetal and maternal anatomy across various sequences and scanning conditions. Development of a machine learning method to effectively address this task requires a large and rich labeled dataset that has not been previously available. Currently, there is no method for accurate fetal brain extraction on various fetal MRI sequences. In this work, we first built a large annotated dataset of approximately 72,000 2D fetal brain MRI images. Our dataset covers the three common MRI sequences including T2-weighted, diffusion-weighted, and functional MRI acquired with different scanners. Moreover, it includes normal and pathological brains. Using this dataset, we developed and validated deep learning methods, by exploiting the power of the U-Net style architectures, the attention mechanism, multi-contrast feature learning, and data augmentation for fast, accurate, and generalizable automatic fetal brain extraction.
 Our approach leverages the rich information from multi-contrast (multi-sequence) fetal MRI data, enabling precise delineation of the fetal brain structures.
 Evaluations on independent test data show that our method achieves accurate brain extraction on heterogeneous test data acquired with different scanners, on pathological brains, and at various gestational stages. 
 This robustness underscores the potential utility of our deep learning model for fetal brain imaging.
 For easy accessibility, we have dockerized our method and made it available on GitHub:\href{https://github.com/bchimagine/fetal-brain-extraction}{https://github.com/bchimagine/fetal-brain-extraction}.
\end{abstract}


\begin{IEEEkeywords}
Deep Learning, Fetal Brain Extraction, Multi-domain Learning, Multi-sequence Fetal MRI
\end{IEEEkeywords}


\section{Introduction}
\label{sec:introduction}
\begin{figure*}[t!]
    \centerline{\includegraphics[width=0.8\textwidth]{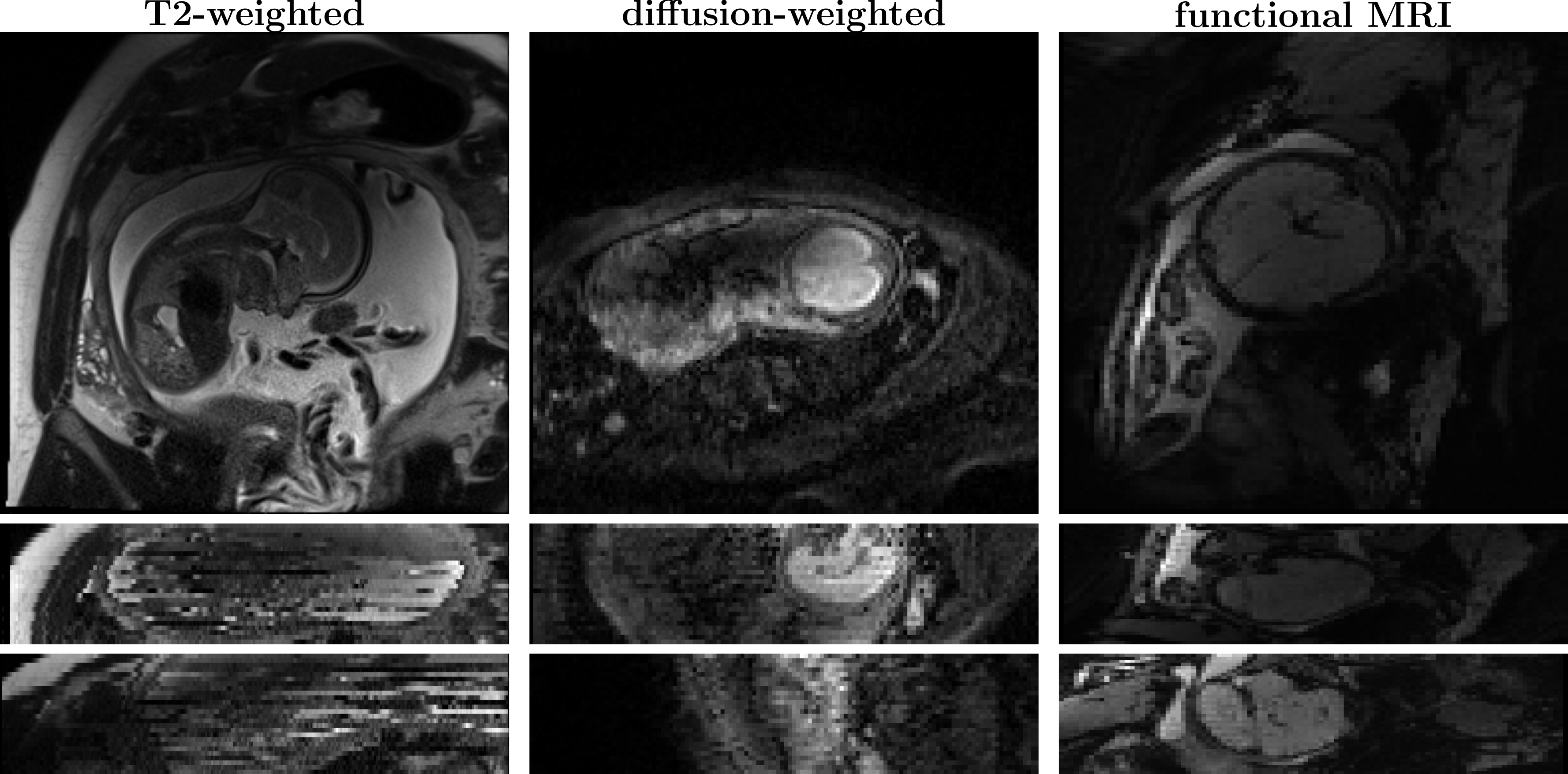}}
    \caption{Examples of multi-modal in-utero MRI images including T2-weighted, diffusion-weighted, and functional MRI. The first row shows in-plane views and the second and the third rows show out-of-plane views. These examples highlight some of the factors that make fetal brain extraction especially challenging such as motion artifacts, anisotropic resolution, heterogeneous contrast, an the highly variable shape and features of the anatomy based on the gestational age and the position of the fetus.}
    \label{fig:fetal_brain}
\end{figure*}

 \IEEEPARstart{F}{etal} Magnetic Resonance Imaging (MRI) is a critical tool in studying prenatal neurodevelopment due to its superior soft tissue contrast compared to ultrasound~\cite{rutherford2008mr}. However, MRI is very susceptible to motion and fetuses can move significantly during MRI scans. To mitigate this problem, fast MRI acquisition techniques are used to obtain stacks of 2D slices.  Brain extraction in these MRI slices is a fundamental step in various applications, including slice-level motion correction and slice-to-volume reconstruction~\cite{gholipour2010robust, kuklisova2012reconstruction, kainz2015fast, ebner2018automated, uus2020deformable, rousseau2006registration, faghihpirayesh2023automatic}. However, automated fetal brain extraction remains challenging due to the variability in brain size, shape, and structure across gestational age, unpredictable fetal and maternal motion, image distortions, intensity non-uniformity, and changing contrast in fetal MRI.
 As can be seen in the examples presented in Fig.~\ref{fig:fetal_brain}, motion artifacts can significantly degrade the quality of fetal MRI acquisitions. Fetal MRI scans also typically exhibit anisotropic resolutions across different axes, featuring high in-plane resolution but lower inter-slice resolution. Consequently, achieving accurate fetal brain extraction from such MR images poses a significant challenge.

 Over the years, various approaches have been proposed to tackle the challenging task of fetal brain extraction, ranging from classical image processing techniques to modern machine learning-based techniques.
 Classical techniques often rely on thresholding, region growing, and morphological operations for fetal brain segmentation~\cite{ison2012fully}. Although these methods may yield acceptable outcomes in specific scenarios, they often  face difficulties related to intensity variations, image artifacts, and the intricate anatomical structures inherent to fetal brain imaging.
 Classical machine learning techniques have also been explored for fetal brain extraction in MRI~\cite{keraudren2014automated}. However, these methods exhibit limitations in terms of efficiency, accuracy, and their ability to perform well in diverse settings. Typically, they follow a two-stage process: first, a technique is applied to detect the fetal brain or a reference object within the MRI, and then the extraction phase isolates the fetal brain from the identified region. One drawback of these approaches is their reliance on manually crafted features to interpret MR images, which may not adequately capture the intricate visual patterns present in the fetal brain and its surrounding tissues~\cite{sun2023multi}.
 The emergence of deep learning has brought significant improvements in the field of medical imaging~\cite{shen2017deep}, including fetal brain extraction~\cite{salehi2018real, rutherford2022automated}. These techniques have demonstrated remarkable performance in handling the complexity and variability inherent in fetal brain MRI scans. 

 While significant strides have been made in addressing the complexities of fetal brain extraction through various methodologies, the quest for more robust, efficient, and accurate techniques persists. The focus of this study is to leverage the power of deep learning to further advance the field of fetal brain MRI extraction, particularly across diverse MRI sequences including T2-weighted (T2W), diffusion-weighted (DWI), and functional MRI (fMRI).

 In this work, we propose a deep learning framework to tackle the challenges posed by fetal brain MRI scans. By capitalizing on the vast information encoded in multi-sequence fetal MRI data, our method aims to surpass the limitations of conventional techniques and classical machine learning approaches.
This work has two primary objectives. First, to achieve superior accuracy in fetal brain extraction. Our goal is to develop methods that work well in the presence of motion artifacts, intensity variations, and complex anatomical structures observed in T2w, DWI, and fMRI sequences. Second, to provide a standardized, efficient, and adaptable solution for seamless integration into the workflow of clinical practitioners and researchers working with diverse MRI modalities. To the best of our knowledge, there is currently no model that can effectively extract fetal brain from MRI images of varying contrasts from different sequences. 

 In the following sections, we will review related research on fetal brain MRI extraction, explain our framework's design and implementation, present the results of our model's performance, and discuss the broader implications and future research directions.


\section{Related Work}
\label{sec:RelatedWork}
 Brain extraction refers to computational methods for removing the skull and and other non-cerebral tissues from head scans. It is an important step that can significantly impact all downstream processing and analyses. In the context of fetal MRI brain extraction, 2D segmentation techniques are often preferred over 3D approaches as acquisitions are based on 2D sequences to minimize through-slice motion artifacts. These acquisitions are often anisotropic in resolution (with relatively thick slices) and almost always exhibit significant inter-slice motion artifacts. Such artifacts as well as maternal respiratory dynamics can result in errors when 3D segmentation methods are used~\cite{sun2023multi}. 
 
 In the case of 2D fetal MRI scans, the fetal brain must be recognized and separated in each slice from a widely variable set of structures such as the uterus, placenta, amniotic fluid, maternal tissue and organs, or fetal body and extremities. In 2D views many of these structures may resemble a sectional view of a developing fetal brain. Therefore, fetal brain extraction on the original MRI slices is a challenging task. Although several studies have addressed fetal brain extraction on 3D reconstructed fetal brain MRI images~\cite{anquez2009automatic, salehi2017auto, khalili2017automatic, ebner2018automated}, only a few have tackled the more difficult task of extracting the fetal brain in each slice of the original acquisitions~\cite{salehi2018real, faghihpirayesh2022deep, li2020automatic, sun2023multi}. Additionally, most existing models for fetal brain extraction have focused on a single sequence, often T2W MRI, which limits their applicability to other MRI sequences.
 
 To address the fetal brain extraction task, several techniques employ a preliminary step of localizing the fetal brain to direct brain extraction. Subsequently, two strategies are presented: either extracting the fetal brain from the entire field of view or after a brain localization step.
 Keraudren et al.~\cite{keraudren2014automated} pinpointed the fetal brain region using a Bag-of-Words model combined with SIFT features~\cite{lowe2004distinctive}. Subsequently, a combination of a sparse patch-based method and a conditional random field was used for brain segmentation from 2D MRI slices. However, these approaches relied on handcrafted features, which often introduced inaccuracies in fetal brain extraction due to the inherent heterogeneity between these features and the subsequent extraction algorithms~\cite{sun2023multi}.

 Taimouri et al.~\cite{taimouri2015template} proposed a block matching technique to simultaneously detect a bounding box around the brain and match the orientation of the brain to a template. Due to a search in the space of possible orientations, this technique was also computationally expensive.
 Tourbier et al.~\cite{tourbier2015automatic} proposed an atlas-based fetal brain segmentation approach that required a predefined bounding box around the brain. This method was also time consuming as it relied on deformable registration to multiple atlases.

 Recent studies have focused mainly on deep learning (DL) and, in particular, convolutional neural networks (CNNs). In comparison to conventional methods, deep CNNs can be utilized to learn the features of fetal brain MR images that are pertinent to the task of fetal brain extraction. The U-Net~\cite{ronneberger2015u} style architecture is the baseline model that is commonly used for medical image segmentation~\cite{ciceri2023review, isensee2021nnu}.


 
 Lou et al.~\cite{lou2019automatic} presented a multistage 2D U-Net, DS U-Net, for fetal brain extraction in T2W MRI. The approach uses a three-step process, all employing the U-Net architecture~\cite{ronneberger2015u}. First, a coarse segmentation is performed to outline a 3D bounding box around the fetal brain. Next, a more detailed segmentation is carried out for precise brain extraction. Finally, a refined segmentation is conducted using a local patch strategy. 
 Li et al.~\cite{li2020automatic} designed a two-step framework using two 2D FCNs~\cite{long2015fully} for fetal brain extraction from MRI slices. One FCN locates and extracts the region of interest (ROI) containing the brain, while a deeper FCN further refines the segmentation. 
 Dudovitch et al.~\cite{dudovitch2020deep} introduced a DL method with two CNN types: a custom 3D U-Net~\cite{cciccek20163d} for bounding box definition and brain extraction, and a 2D U-Net~\cite{ronneberger2015u} that refines segmentation considering adjacent slice results. 

 Liao et al.~\cite{liao2020joint} showcased a multistage DL model for both image quality assessment and fetal brain extraction, with modules that detect and extract the brain using the U-Net and deformable convolutional layers. 
 Zhang et al.~\cite{zhang2021confidence} proposed a confidence-aware cascaded framework with two U-Net~\cite{ronneberger2015u} modules, one for localization and another for fine-tuning. The framework evaluates slice-specific confidence for extraction, using higher-confidence slices to guide the extraction of lower-confidence ones.
 Salehi et al.~\cite{salehi2018real} utilized a 2D U-Net~\cite{ronneberger2015u} to efficiently extract the fetal brain from T2W MR images.
 Khalili et al.~\cite{khalili2017automatic} presented a multiscale CNN, influenced by the architecture from~\cite{moeskops2016automatic}, using three parallel 2D convolutional pathways that analyze 2D patches of different sizes. In a subsequent study, Khalili et al.~\cite{khalili2019automatic} utilized a 2D CNN based on a scaled-down U-Net architecture~\cite{ronneberger2015u} for both fetal and neonatal T2W MR scans, and a post-processing algorithm from their previous work~\cite{khalili2017automatic}. 


 Faghihpirayesh et al.~\cite{faghihpirayesh2022deep} proposed RFBSNet, a U-Net style~\cite{ronneberger2015u} architecture designed for real-time fetal brain segmentation, emphasizing speed and accuracy.
 Rutherford et al.~\cite{rutherford2022automated} introduced a CNN model adapted from the U-Net architecture specifically for fetal brain extraction in fMRI. 

 While all of the above-referenced studies focused on only one type of MRI modality, either T2W or fMRI; in this study we aimed to train a model on a pool of heterogeneous data from diverse MRI modalities including T2W, fMRI, and DWI. We anticipated that this approach should improve the model's performance by 1) enabling the model to recognize features that are specific to each MRI sequence, and 2) allowing the model to learn patterns that are common to all modalities. We hypothesized that this holistic training approach would help improve model robustness in handling varying imaging scenarios and enhance generalizability in real-world settings.
 
 To build a holistic fetal brain extraction tool for fetal MRI (Fetal-BET), we built deep learning models based on some of the best-performing CNN architectures. Specifically, the U-Net, the nnU-Net (here the dynamic U-Net), and the Attention U-Net. We critically evaluated the performance of these models on our heterogeneous test sets. Since none of the previous methods were designed to segment the brain on multi-contrast images, a direct comparison was not appropriate. Nonetheless, we critically evaluated the performance of Fetal-BET on any specific image type, where we specifically evaluated the performance gain achieved by multi-sequence feature learning and data augmentation. We built and trained models based on variations of the U-Net that was used in most of the previous studies~\cite{salehi2018real,lou2019automatic,liao2020joint, dudovitch2020deep,rutherford2022automated,faghihpirayesh2022deep}.

\section{Materials and Methods}
\label{sec:Method}
\subsection{Data}
\label{sec:Data}
 The data utilized in this study were sourced from fetal MRIs conducted at Boston Children’s Hospital over a span of approximately 20 years. These MRI acquisitions encompassed a range of MRI scanners, including 1.5T GE, Philips, and Siemens, as well as 3T Siemens scanners, specifically Skyra, Prisma, and Vida models. The study was approved by the Institutional Review Board. For all prospective fetal MRIs, written informed consent was obtained from the participants.

 The MRI acquisition protocols typically involved the acquisition of multiple types of images, including T2-weighted (T2W) 2D half-Fourier single-shot turbo spin echo sequences with in-plane resolutions ranging from 1 to 1.25 mm and slice thicknesses between 2 to 4 mm, capturing detailed structural information. Additionally, diffusion-weighted imaging (DWI) was acquired with an in-plane resolution of 2 mm and slice thickness ranging from 2 to 4 mm, enabling the assessment of water diffusion in fetal brain tissues. Functional MRI (fMRI) images were also included, featuring an isotropic resolution of 2-3 mm, allowing the study of brain activity and neural circuit activity development in the developing fetal brain.

 Our dataset included a total of 38,038 2D MRI slices (from 100 subjects) for T2W imaging, 22,902 2D MRI slices (from 65 subjects) for DWI, and 4756 2D MRI slices (from 36 subjects) for fMRI. The fetal scans in this dataset span a wide gestational age range, ranging from 22 to 38 weeks, resulting in considerable variations in brain size and shape—approximately a five-fold increase in brain volume over this period. Moreover, the dataset exhibits diversity by encompassing a spectrum of conditions, including both typical and abnormal brains, various artifacts, and twin pregnancies. This diversity mirrors the complexities encountered in real-world fetal MRI imaging, facilitating robust evaluation and analysis of the developed algorithms.
 For ground truth annotations, a skilled annotator meticulously segmented the fetal brain on each MRI slice. 

 To ensure proper evaluation, we partitioned the data into three subsets: training, validation, and testing. This partitioning was performed on a subject-wise basis such that there was no overlap between subjects in different subsets. This was necessary to ensure independent evaluation of our model's performance. Additionally, we augmented our test dataset with a separate collection of fetal MRI scans sourced from different scanners and distinct imaging sites. This supplementary dataset was entirely excluded from the training phase, ensuring its independence from our model development process. This was critical to test the generalization performance of our proposed methods to unseen data from different sites and scanners.
 For a comprehensive overview of the data distribution across these subsets, please refer to Table~\ref{table:data}.


\begin{table}[t!]
    \caption{Summary of the fetal MRI data used in this work. We used three different MRI sequences including T2-weighted (T2W), diffusion-weighted MRI (DWI), and functional MRI (fMRI). T2W/Typical refers to most common seen data types in fetal T2W MRIs including normal and challenging cases. T2W/SiteW involved independent T2W scans acquired at a remote site. DWI/B0 represents the non-diffusion sensitized baseline images in DWI, while DWI/B1 refers to the collection of diffusion-sensitized images. fMRI/External refers to independent fMRI data that we accessed through OpenNeuro~\cite{rutherford2022automated}.}
    \label{table:data}
    \begin{adjustbox}{width=\columnwidth,center}
    \centering
    \renewcommand{\arraystretch}{1.5}
    \begin{tabular}{llll}
    \hline
    Modality/Type & Subjects (Stacks, Slices) & \makecell{Resolution (mm)} & Scanner\\
    \hline

    Training & & & \\
    \hline
    T2W & 44 (483, 19083) & $\sim1\times1\times2$ & Siemens3T \\
    DWI & 41 (492, 14597) & $\sim2\times2\times2-4$ & Siemens3T \\
    fMRI & 22 (85, 2790) & $\sim2-3$ isotropic & Siemens3T \\
    \hline
    
    Validation & & & \\
    \hline
    T2W & 9 (82, 3236) & $\sim1\times1\times2$ & Siemens3T \\
    DWI & 9 (108, 3260) & $\sim2\times2\times2-4$ & Siemens3T \\
    fMRI & 5 (25, 750) & $\sim2-3$ isotropic & Siemens3T \\
    \hline
    
    Testing & & & \\
    \hline
    T2W/Typical & 18 (143, 5362) & $\sim1\times1\times2$ & Siemens3T \\
    T2W/Abnormality & 4 (30, 1124) & $\sim1\times1\times2$ & Siemens3T \\
    T2W/Artifacts & 6 (34, 1238) & $\sim1\times1\times2$ & Siemens3T \\
    T2W/Twins & 3 (32, 922) & $\sim1\times1\times2$ & Siemens3T \\
    T2W & 3 (14, 391) & $\sim1\times1\times2$ & GE1.5T \\
    T2W & 5 (84, 2732) & $\sim1\times1\times2$ & Phillips1.5T \\
    T2W & 3 (28, 947) & $\sim1\times1\times2$ & Siemens1.5T \\
    T2W/SiteW & 5 (91, 3003) & $\sim1\times1\times2$ & Siemens3T \\
    DWI/B0 & 14 (46, 1312) & $\sim2\times2\times2-4$ & Siemens3T \\
    DWI/B1 & 14 (122, 3733) & $\sim2\times2\times2-4$ & Siemens3T \\
    fMRI & 9 (38, 1216) & $\sim2-3$ isotropic & Siemens3T \\
    fMRI/External & 77 (477, 17649) & $\sim3$ isotropic & Siemens3T \\ 
    \hline
    \end{tabular}
    \end{adjustbox}
\end{table}


\subsection{Model Architecture}
\label{sec:Model}
 In our study, we investigated three state of the art neural network architectures for medical image segmentation:
 U-Net~\cite{ronneberger2015u}, Dynamic U-Net~\cite{ranzini2021monaifbs,futrega2021optimized}, which is an adaptation of the nnU-Net framework~\cite{isensee2021nnu}, and Attention U-Net~\cite{oktay2018attention}.

\begin{figure*}[t!]
    \centerline{\includegraphics[width=1.0\textwidth]{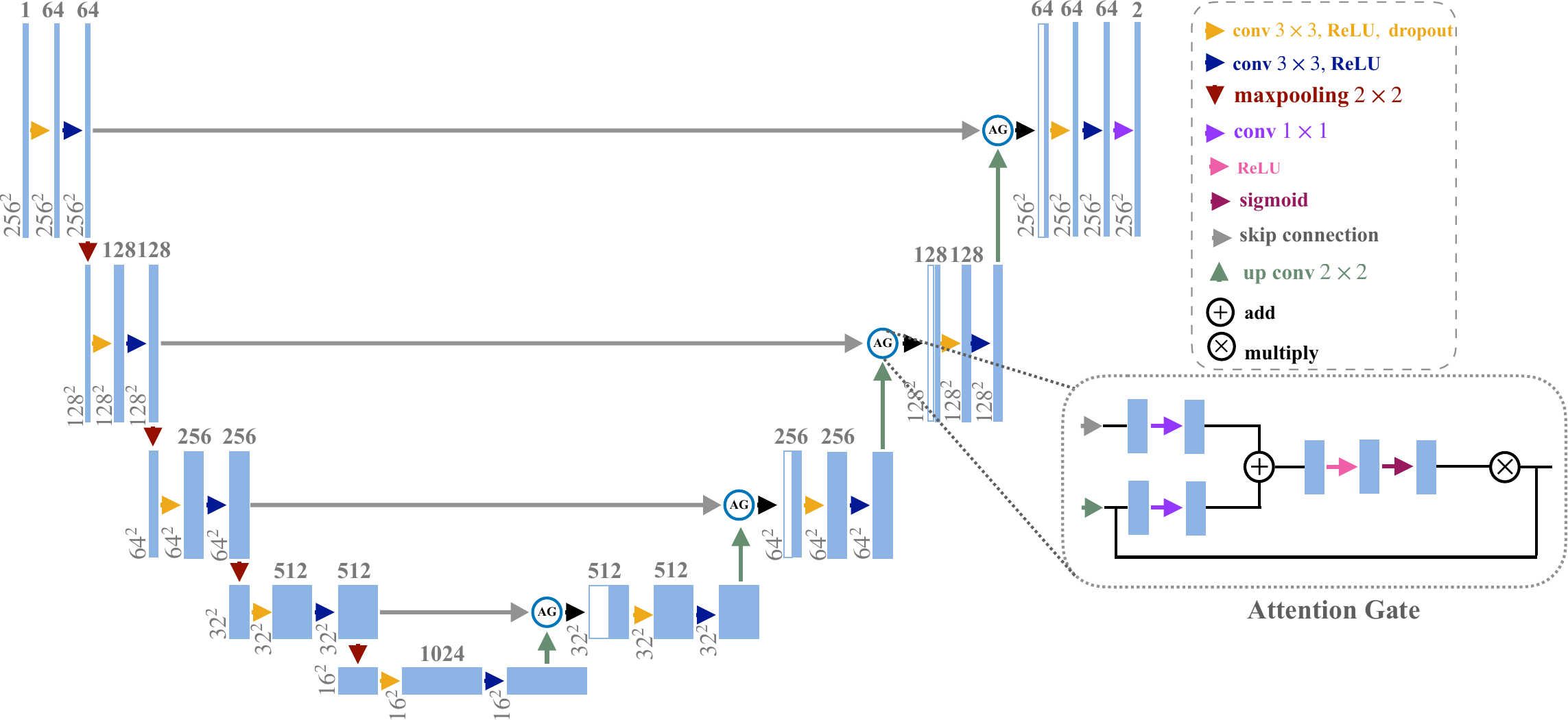}}
    \caption{Architecture of the U-Net with Attention Gates (AG) known as Attention U-Net. The backbone U-Net architecture can be achieved if the AG units are ignored. In the Attention U-Net, AGs filter the features that are propagated through the skip connections by using the contextual information of features extracted in coarser scales. This is achieved by adding the decoder output of a coarser scale to the output of every skip connection from the encoder after $1 \times 1$ convolutions. The output then passes through ReLU and sigmoid activation functions and is multiplied to the coarser level decoder input.}
    \label{fig:net}
\end{figure*}

 U-Net~\cite{ronneberger2015u} is a standard benchmark in the field of biomedical image segmentation.
 It features a symmetrical encoder-decoder structure, where the encoder extracts hierarchical features from the input data, and the decoder progressively upsamples and refines these features to produce the final segmentation map. U-Net is widely recognized for its adaptability through skip connections that concatenate feature maps from the encoder to the decoder, allowing the model to combine multi-scale contextual information with fine-grained details. The adaptability of U-Net lies in its ability to capture both local and global context effectively. However, it may struggle with handling intricate anatomical structures and fine details in certain scenarios due to its fixed architecture and limited contextual awareness. In Fig.~\ref{fig:net}, we illustrate an instance of a U-Net architecture without attention gates, showcasing its standard structure for segmentation tasks.
 
 Dynamic U-Net~\cite{futrega2021optimized, ranzini2021monaifbs} is an adaptation of the nnU-Net framework~\cite{isensee2021nnu}. It showcases its capacity to autonomously customize its architecture according to the input data. Notably, Dynamic U-Net demonstrates intelligent adaptability in selecting optimal parameters such as kernel sizes, strides, and channel dimensions, all tailored to the specific input image size. This flexibility is expected to enhance network efficiency and its ability to seamlessly adapt to diverse datasets.

 Attention U-Net~\cite{oktay2018attention} enhances the base U-Net by introducing an attention gate (depicted in Fig.~\ref{fig:net}) within the decoder section. The attention gate processes the encoder's feature map before concatenation in the decoder block, determining the significance of regions in the encoder feature map relative to the context of the preceding decoder block. This determination is facilitated by the multiplication of the encoder feature map with attention gate-computed weights, which range between 0 and 1, representing the neural network's focus on specific pixels. The attention mechanism essentially acts as a gatekeeper, facilitating the model's ability to attend to the most salient features, ultimately improving segmentation accuracy by adaptively selecting and combining information from various spatial locations. Fig.~\ref{fig:net} presents a detailed visual representation of the Attention U-Net structure, highlighting the integrated attention modules and demonstrating how they interact with the traditional U-Net (bypassing Attention Gate) layers to achieve more discerning and accurate segmentations.


\subsection{Model Training and Inference}
\label{sec:implementation}
 
 \subsubsection{Data processing and augmentation}
 \label{sec:Pre-processing}
 
 First, to establish uniformity across all three modalities (T2W, DWI, and fMRI) for training (only), we resampled each slice in two dimensions to a consistent pixel size of 1mm and resized them to a uniform image size of $256 \times 256$ pixels. This ensured that all preprocessed slice images adhered to the same dimensions, denoted as (H, W, C), with H=W=256 representing width and height, and C=1 signifying the number of image channels. 
 Furthermore, the data is normalized on a per-slice basis by its variance, ensuring uniform data scaling and facilitating improved model convergence.

 Second, to bolster the robustness of our method against challenges that are inherent in fetal MRI, we used a comprehensive set of data augmentation techniques. We designed these techniques specifically to tackle the challenges that are raised by the widely variable position, orientation, size, and shape of the fetal brain and its surrounding structures, as well as the pronounced effects of fetal movements and maternal breathing motion. Depending on the sequence type, intermittent fetal and maternal motion often result in pronounced artifacts that may include partial or complete signal loss, in-plane blur, ringing, slice cross-talk, and spin-history artifacts.
 
 In our image augmentation pipeline, we applied a range of transformations to enhance the diversity of our training dataset. Specifically, we generated augmented scans using the following techniques: 
 
 1. Spatial Augmentation: This category consisted of five spatial transformations, including random flips, rotations, zooming, and affine transformations. For each augmentation technique, we applied these transformations with a probability of 0.5 or 0.6, resulting in a significant increase in the variability of the dataset.
 
 2. Intensity Augmentation: In this category, we introduced variations in image intensity through three distinct techniques. Gaussian noise was added with a standard deviation of 0.4 and a probability of 0.5. Multiplicative bias fields were incorporated with a degree of 4 and coefficients ranging between 0.05 and 0.1, applied with a probability of 0.6. Gaussian smoothing was performed with sigma values ranging from 0.5 to 1.0 and a probability of 0.4.

 These augmentations collectively contributed to a versatile training dataset, allowing our deep learning models to better adapt to diverse fetal brain MRI images during training.

 \subsubsection{Training}
 \label{sec:Training}
 In the training stage, we initiated the process by using an independent validation dataset to determine the best hyperparameters. Subsequently, we trained the models with the training dataset. All models were trained with a batch size of 8 and input image size fixed at $256 \times 256$. The learning rate for each of the compared networks was fine-tuned separately. For the Attention U-Net, we set a learning rate of $1 \times 10^{-4}$. We conducted training for 300 epochs using Adam optimization~\cite{kingma2014adam} of stochastic gradient descent.

 To guide the training process, we utilized a weighted sum of binary cross-entropy and Dice loss. Specifically, we employed the batched variant of the Dice loss, where the loss was calculated over all samples in the batch, as opposed to averaging the Dice loss for each 2D sample individually. This approach considers samples in the batch as a pseudo-volume and computes the loss function over all voxels in the batch~\cite{isensee2021nnu}. This was done for improved computations only, as models only worked on 2D slices. Since the original fetal MRI acquisitions were 2D (slice-based), to minimize the effects of inter-slice motion, no 3D context was used in training, inference, or evaluation. We used foreground Dice, which measures the overlap between the predicted and ground truth segmentations. Our loss function was
 \begin{equation}
     L_{total} = 0.4 \times L_{CE} + 0.6 \times L_{dice}
 \end{equation}
 \begin{equation}
      L_{CE} = -\sum_{i=1}^{n} v_i \log(u_i) + (1-v_i) \log(1-u_i)
 \end{equation}
 \begin{equation}
     L_{dice} (u, v)= -\frac{2}{|K|}\sum_{k\in K}\frac{\sum_{i\in I}u_i^kv_i^k}{\sum_{i\in I}u_i^k+\sum_{i\in I}v_i^k}
 \end{equation}
 where $u$ is the softmax output of the network and $v$ is a one hot encoding of the ground truth segmentation map. Both $u$ and $v$ have shape $I \times K$ with $i \in I$ being the number of pixels in the training batch and $k\in K$ being the classes.

 \subsubsection{Inference}
 \label{sec:Inference}
 During the inference stage, we apply the same resampling and normalization procedures as in the training data (see section:~\ref{sec:Pre-processing}), ensuring consistency in data preparation. However, unlike the training dataset, we do not perform resizing on these datasets. We utilize a sliding window approach with a window size matching the training patch size ($256\times 256$), allowing flexibility in input image sizes. Each consecutive window overlaps the previous one by half its size, and predictions within these overlapping regions are averaged to produce the final prediction. This method ensures consistent and accurate results across varying input sizes.

\subsubsection{Implementation}
 \label{sec:Implementation}
All our experiments and model training were conducted with NVIDIA RTX A5000 GPUs on a workstation with 128GB of system memory. Our implementations leveraged the PyTorch framework~\cite{paszke2019pytorch} and harnessed the capabilities of the MONAI toolkit~\cite{cardoso2022monai}.

 \subsubsection{Evaluation}
 \label{sec:Evaluation}
 We evaluated the performance of all methods using two key metrics: the Dice Similarity Coefficient (DSC) and Intersection-over-Union (IoU), also known as the Jaccard index. The DSC, a widely used metric in segmentation tasks, quantifies the overall overlap between two sets of pixels, such as the predicted and ground truth segmentations. In addition to DSC, we employed the IoU metric, which provides additional spatial context by measuring the ratio of the intersection to the union of these sets. DSC and IoU are defined as:
 \begin{equation} 
     DSC(P, R) = \frac{2 |P\cap R| }{ |P| + |R| } = \frac{2TP}{2TP+FP+FN}
 \end{equation}

 \begin{equation}
     IoU(P,R) = \frac{ |P\cap R| }{ |P\cup R| } = \frac{TP}{TP+FP+FN}
 \end{equation}
 where P is the predicted mask, R is the Reference (ground truth) mask. TP, FP, and FN are the true positive, false positive, and false negative rates, respectively.

\begin{table*}[t!]
\caption{Average Dice (Dice), average IoU, and inference time on test data for different models.}
    \centering

        \centering
        
        \label{table:results}
        \resizebox{0.95\textwidth}{!}{%
        \begin{tabular}{l|c|c|c|c|c|c|c}
            \hline
            & \multicolumn{2}{c|}{T2W} & \multicolumn{2}{c|}{DWI} & \multicolumn{2}{c|}{fMRI} & Inference Time \\
            \cline{2-7}
            Data & Dice (\%) & IoU (\%) & Dice (\%) & IoU (\%) & Dice (\%) & IoU (\%) & (ms)  \\
            \hline
            U-Net~\cite{ronneberger2015u} & 95.72$\pm$4.24 & 92.06$\pm$6.93 & 92.58$\pm$5.83 & 86.66$\pm$8.78 & 95.05$\pm$2.27 & 90.66$\pm$3.96 & 8.18$\pm$0.74\\
            DynUNet~\cite{isensee2021nnu} & 93.07$\pm$8.31 & 88.01$\pm$12.56 & 92.36$\pm$4.63 & 86.14$\pm$7.54 & 93.12$\pm$4.18 & 87.40$\pm$6.89 & 10.49$\pm$0.72\\
            AttU-Net~\cite{poudel2019fast} & 95.70$\pm$4.28 & 92.03$\pm$6.97 & 93.27$\pm$4.11 & 87.64$\pm$6.69 & 95.25$\pm$1.77 & 90.99$\pm$3.16 & 11.48$\pm$1.11\\
            \hline
        \end{tabular}%
    }
\end{table*}
    
\begin{figure*}[t!]



        \centering
        \includegraphics[width=\textwidth]{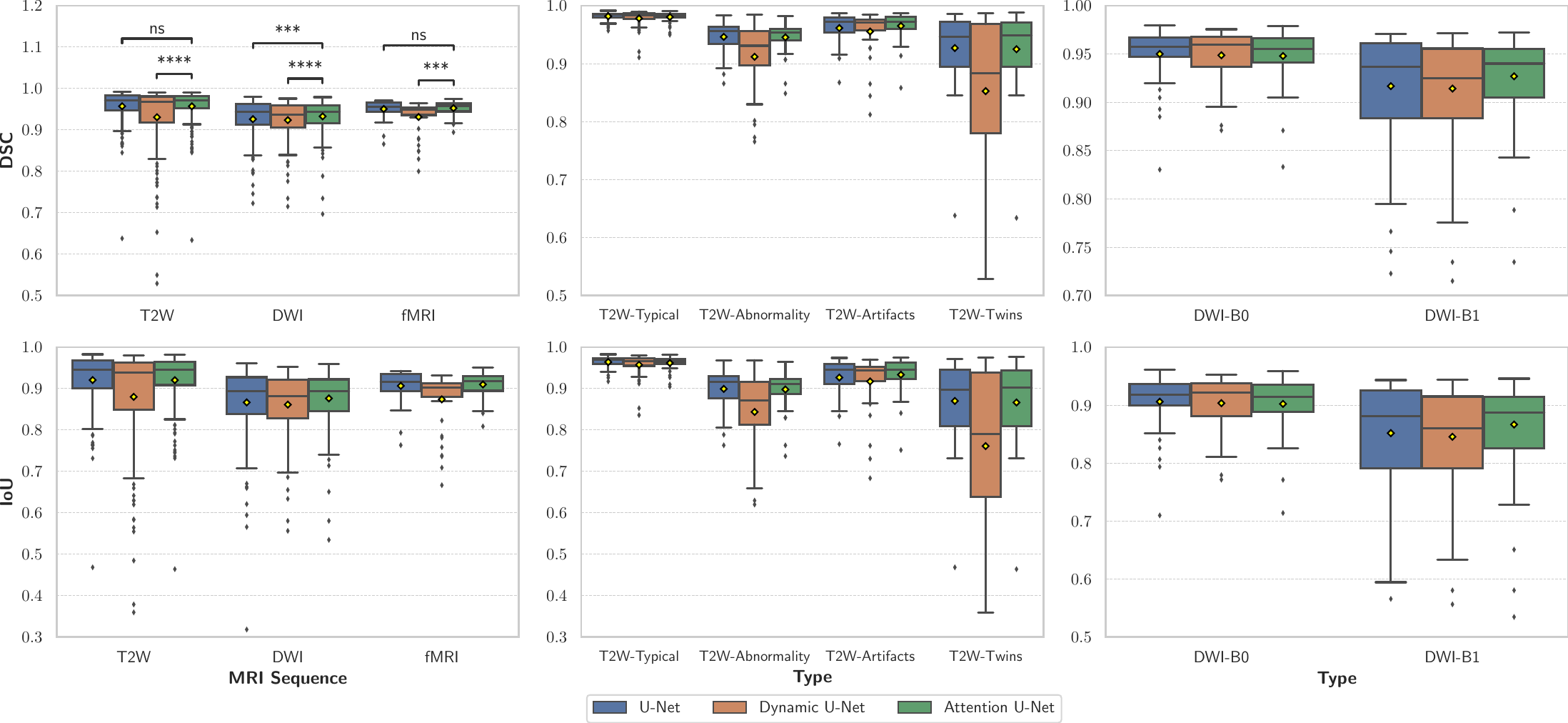}
        \caption{Boxplots of Dice similarity coeffiicent (DSC) and Intersection-over-Union (IoU) for different sequences (T2-weighted (T2W), diffusion-weighted (DWI), and functional MRI (fMRI)), T2W data characteristics (Typical, Abnormalities, Artifacts, and Twins pregnancies), and model architectures (U-Net, Dynamic U-Net, and Attention U-Net). Higher DSC and IoU values indicate greater segmentation accuracy. The U-Net and Attention U-Net models achieved higher median Dice scores overall compared to the Dynamic U-Net model.
        The asterisks displayed on the top left plot serve as visual indicators of the statistical significance associated with the differences observed between the groups using a paired t-test. The asterisks displayed on the first plot serve as visual indicators of the statistical significance associated with the differences observed between the groups. Significance levels: (ns) $p > 0.05$, (*) $0.01 < p \leq 0.05$, (**) $0.001 < p \leq 0.01$, (***) $0.0001 < p \leq 0.001$, (****) $p \leq 0.0001$.}
        \label{fig:boxplots1}

\end{figure*}


 \begin{figure*}[t!]
     \centering
     \includegraphics[width=0.85\textwidth]{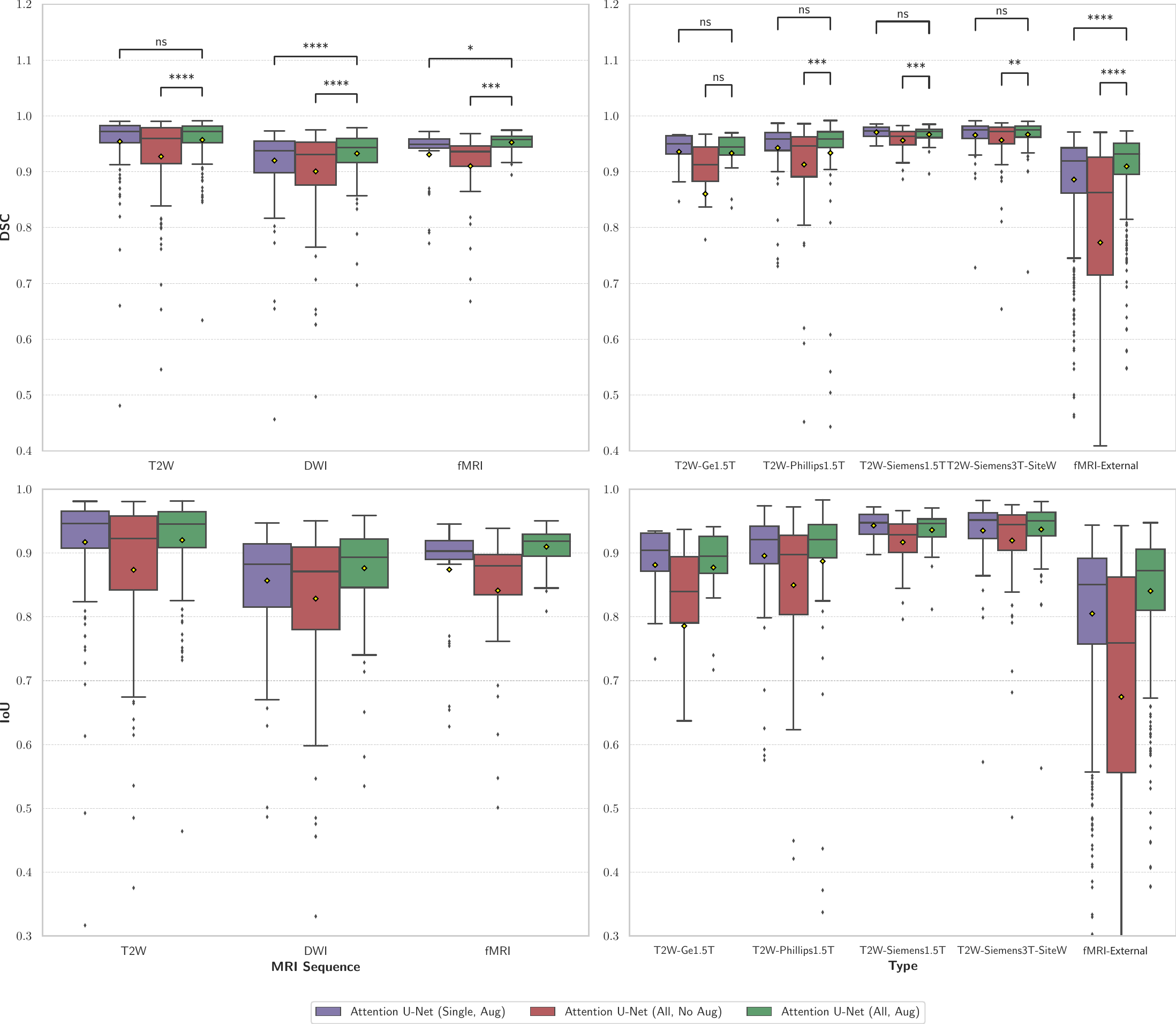}
     \caption{Boxplots of Dice similarity coeffiicent (DSC) and Intersection-over-Union (IoU) for different MRI sequences (T2-weighted (T2W), diffusion-weighted (DWI), and functional MRI (fMRI)), comparing the extraction performance of different Attention U-Net model architectures ((Single, Aug): trained on a single sequence (corresponding) with augmentation, (All, No Aug): trained on all sequences with no augmentation, and (All, Aug): trained on all sequences with augmentation).
     Left plots show the results on our test data and right plots show the results on our out-of-distribution test data.
     The best result is achieved by Attention U-Net trained on all sequences with augmentation. This indicates that leveraging multiple complementary sequences and expanded training data through augmentation can enhance deep learning model performance.
     }
     \label{fig:boxplots_alb}
 \end{figure*}

\begin{figure}[t!]
    \centerline{\includegraphics[width=0.94\columnwidth]{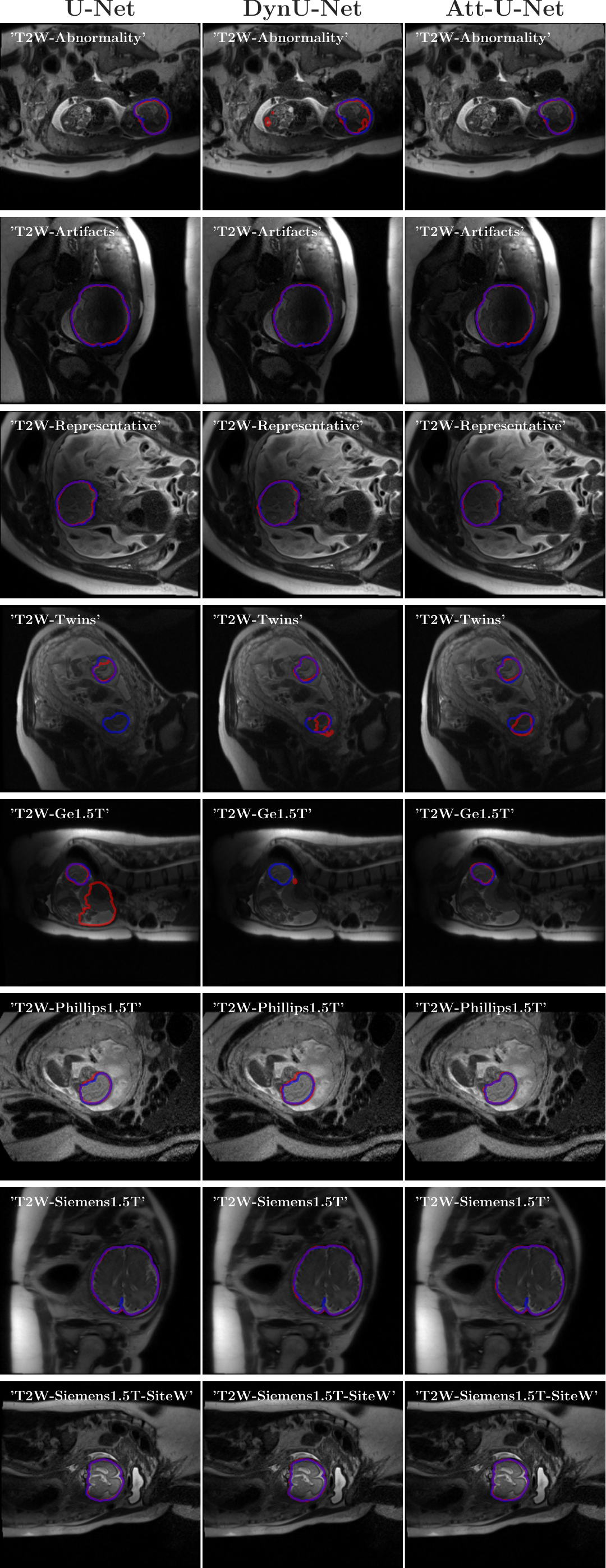}}
    \caption{Representative examples of segmentations produced by different models (U-Net, DynU-Net, AttU-Net) on T2w slices. On each image, the blue curve shows the outline of the reference brain mask drawn manually by an experienced annotator, while the red curve shows the contour of the segmentation mask predicted by the deep learning method.}
    \label{fig:badsliceexa}
\end{figure}


\begin{figure}[t!]
    \centerline{\includegraphics[width=\columnwidth]{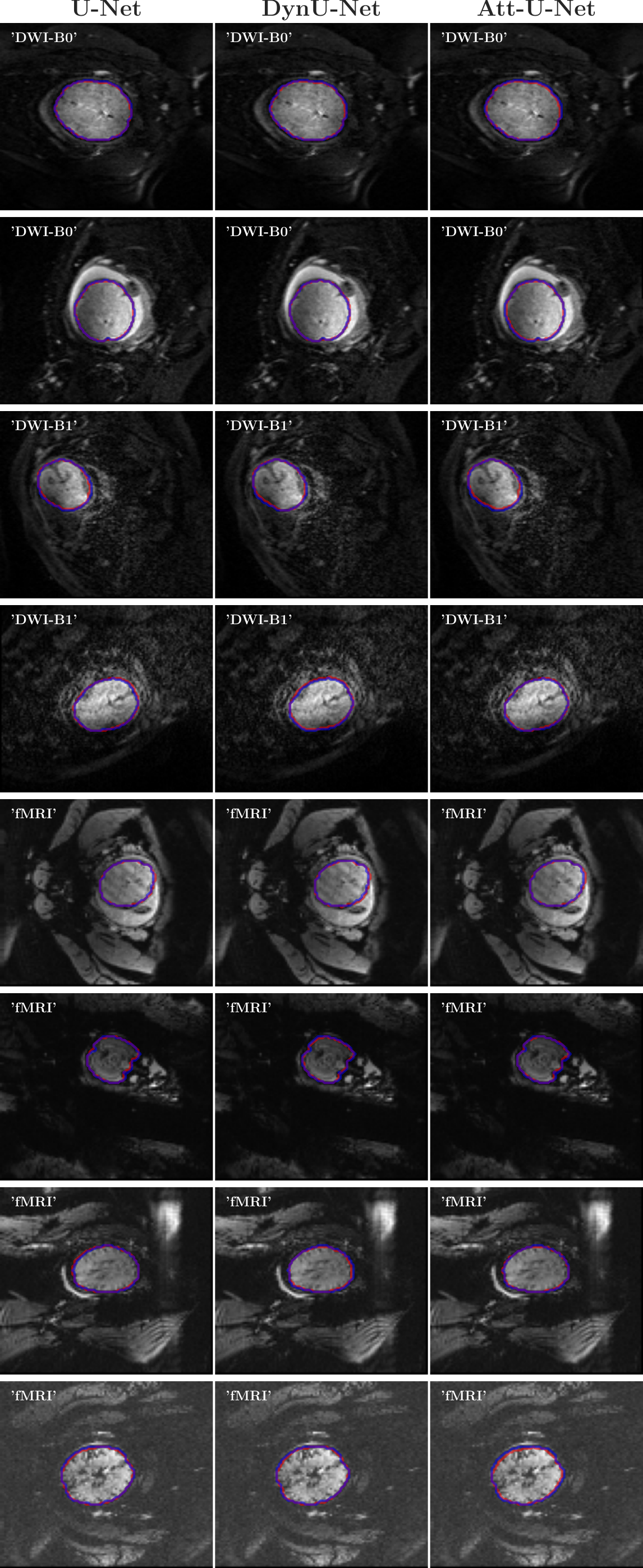}}
    \caption{Representative examples of segmentations produced by different models (U-Net, DynU-Net, AttU-Net) on DWI and fMRI test slices. The blue curves show the outline of the reference (ground truth) brain mask, while the red curve shows the contour of the mask predicted by the deep learning method.}
    \label{fig:badsliceexa2}
\end{figure}

\section{Results}
\label{sec:Results}
 In this section, we first present the results of training the three models, i.e., the U-Net, the dynamic U-Net (DynU-Net), and the attention U-Net (AttU-Net) on all sequences. We compared the performance of the three models on the different test sets that we built.
 Next, we conducted ablation studies to evaluate the relative impact of our proposed multi-sequence learning and data augmentation strategies. To this end, we compared the best model performance with models trained on each individual sequence, as well as models trained without data augmentation. Specifically, we assessed how our best model trained on all sequences performed compared to models trained on each sequence individually. This comparative analysis provides insights into the impact of multi-sequence training and data augmentation, demonstrating how they increase the robustness and generalizability of deep learning models.
 
 Table~\ref{table:results} provides a detailed breakdown of Dice and IoU scores for each of the considered architectures and sequences when evaluated on an independent test dataset. In this experiment, all models were trained on a diverse dataset encompassing all three sequences (T2W, DWI, and fMRI). Overall, the results show that the standard U-Net and Attention U-Net models are very accurate in extracting the fetal brain in all MRI sequences. Notably, U-Net and Attention U-Net achieved a DSC of 95.72\% and 95.70\% and an IoU of 92.06\% and 92.03\% for the T2W sequence respectively, demonstrating their effectiveness in fetal brain extraction on this sequence. For the DWI sequence, Attention U-Net achieved a DSC of 93.27\% and an IoU of 87.64\%, whereas U-Net achieved a DSC of 92.58\% and an IoU of 86.66\%, showcasing Attention U-Net's robustness across diverse MRI sequences. In the case of fMRI, the Attention U-Net model yielded a DSC of 95.25\% and an IoU of 90.99\%, underlining its capability to handle complex functional MRI data. While the dynamic U-Net also delivered commendable results, with slightly lower DSC and IoU scores, this could be attributed to its inherent sensitivity to dynamic changes in the input data. The dynamic U-Net adapts its architecture to the input size dynamically, which might lead to minor fluctuations in performance~\cite{ranzini2021monaifbs}.

 Fig.~\ref{fig:boxplots1} shows the boxplots that provide an illustration of the distribution and variability of the DSC and IoU scores across the different test datasets for each of the trained models. The first column of boxplots in Fig.~\ref{fig:boxplots1} represents results obtained on all T2W, DWI, and fMRI test data. The second column offers a detailed analysis of the performance of the three models on the T2W images including typical fetal brains, abnormal cases, images with artifacts, and twin pregnancies. The boxplots provide insights into the adaptability of the trained models to varying T2W data scenarios. In the third column, we focused on the evaluation of the DWI sequences, specifically considering B0 and B1 data.

 We used paired t-tests to assess the statistical significance of the difference between the performance of different models. For comparing U-Net with Attention U-Net, the p-value for T2W, DWI, and fMRI were, respectively, \(8.585e-01 \), \(2.643e-04 \), and \(2.250e-01 \). For comparing Dynamic U-Net and Attention U-Net, the p-values were \(2.957e-08 \), \(5.301e-06 \), and \(8.494e-04 \), respectively. The asterisks displayed on the top left plot serve as visual indicators of the statistical significance associated with the differences observed between the groups.
 (ns) indicate no significant difference (p-value $> 0.05$).
 ($*$) suggests moderate significance ($0.01 <$ p-value $\leq 0.05$).
 ($**$) signify high significance ($0.001 <$ p-value $\leq 0.01$).
 ($***$) represent very high significance ($0.0001 <$ p-value $\leq 0.001$).
 ($****$) indicate extremely high significance (p-value $\leq 0.0001$).

 In the ablation studies we focused on gauging the effectiveness of 1) our models trained on multiple sequences when compared to models trained on each sequence separately, and 2) our image augmentation strategies.  The DSC and IoU plots on the left side of Fig.\ref{fig:boxplots_alb} illustrate the results of our ablation studies on the test sets of every sequence (T2W, DWI, fMRI). Each box plot represents the performance distribution of models trained on individual sequences (Attention U-Net, Single Aug), a model trained on all sequences but without data augmentation (Attention U-Net, All, No Aug), and our final model (Attention U-Net, All, Aug), which used all the sequences along with our data augmentation for training. The results show that our final model, Attention U-Net (All, Aug), performed significantly better than Attention U-Net without data augmentation (All, No Aug) for all of the sequences, and performed significantly better than Attention U-Net (Single, Aug) for the DWI and fMRI sequences. 
 
 The DSC and IoU plots on the right side of Fig.\ref{fig:boxplots_alb} show the results on set-aside test sets from different scanners. Since no images from those scanners or sites were included in the training dataset, this served as a test of the generalization performance of the models for data from new domains. Overall, the results show that our final model that exploited multi-sequence learning and data augmentation, performed significantly better than its counterparts that did not use multi-sequence learning or data augmentation. The results in Fig.\ref{fig:boxplots_alb} vividly display how the aggregated model outperformed its counterparts across various test datasets, underscoring the advantages of multi-sequence training and data augmentation to improve generalization.

Fig.\ref{fig:badsliceexa} and Fig.\ref{fig:badsliceexa2} provide sample representative outcomes of our trained model (Fetal-BET) on a variety of test data including images of twins and brains with abnormalities on T2w images as well as DWI and fMRI scans.


\section{Discussion}

 Our findings underscore the effectiveness of our approach in fetal brain extraction. In particular, our experiments with three powerful deep convolutional neural network architectures trained with multiple sequences and comprehensive data augmentation strategies demonstrate that we can achieve accurate automatic fetal brain extraction. Both U-Net and the Attention U-Net models exhibited high DSC and IoU scores, particularly on T2W images, with Attention U-Net showcasing robust generalizability across various MRI contrasts and outperforming the other two models on DWI and fMRI.

 Importantly, our ablation study results, as depicted in Figure~\ref{fig:boxplots_alb}, highlight the advantage of multi-sequence training and data augmentation. The model trained on the amalgamation of all three sequences consistently outperformed models trained individually on each sequence when tested on that specific sequence alone. In several instances, this difference was statistically significant, particularly in achieving higher DSC scores for the more challenging sequences, i.e., DWI and fMRI, when compared to single-sequence models.

 In summary, our study addresses the complex task of fetal brain extraction in fetal MRI analysis. We have developed an innovative solution using deep learning techniques, attention mechanisms, multi-contrast learning, and data augmentation. Crucially, we have created a substantial and diverse dataset that includes various MRI sequences and pathological cases, significantly contributing to the advancement of our approach. Through rigorous evaluation, we demonstrated the reliability and robustness of our method, achieving accurate fetal brain extraction across different scanning conditions, stages of pregnancy, and brain conditions. The adaptability and precision of our deep learning model hold significant promise for the field of fetal brain imaging and analysis. By overcoming long-standing challenges in fetal brain extraction, our work has the potential to improve automatic workflows for quantitative fetal MRI analysis, including, for example, image reconstruction and segmentation. Therefore, it has the potential to profoundly impact clinical practices and advance our understanding of prenatal brain development and developmental disorders, ultimately improving the quality of prenatal care and diagnostics.


\section{Conclusion}
 In conclusion, our study presents a comprehensive evaluation of fetal brain extraction techniques using a range of MRI sequences. We have demonstrated the efficacy of both U-Net and Attention U-Net architectures, with Attention U-Net excelling, particularly in challenging sequences like DWI and fMRI. The advantages of multi-sequence and augmentation training were clearly evident, with the aggregated model consistently outperforming individual sequence models.
 As we move forward, further research may explore fine-tuning these models for specific clinical applications and expanding the dataset to encompass even more variations. The promising results obtained in this study provide a strong foundation for future advancements in the field of fetal brain extraction, ultimately benefiting both healthcare professionals and expectant parents in ensuring the well-being of unborn children.

\section*{Acknowledgment}
 This research was supported in part by the National Institutes of Health (NIH) under award numbers R01NS106030, R01EB018988, R01EB031849, R01-EB032366, R01HD109395, R01NS128281, R01HD110772; and in part by the Office of the Director of the NIH under award number S10OD025111. This research was also supported in part by NVIDIA Corporation and utilized NVIDIA RTX A5000 GPUs. The content of this publication is solely the responsibility of the authors and does not necessarily represent the official views of the NIH or NVIDIA. 


\bibliographystyle{IEEEtran}
\bibliography{references}

\end{document}